\documentclass[aps,prl,twocolumn,superscriptaddress,10pt, showpacs]{revtex4-2}

\usepackage{graphicx}% Include figure files
\usepackage{dcolumn}% Align table columns on decimal point
\usepackage{amsmath}
\usepackage{nicefrac}
\usepackage{upgreek}
\usepackage{epstopdf}
\usepackage{color}
\usepackage[table]{xcolor}
\usepackage{multirow}
\definecolor{magd}{RGB}{154,12,70}
\definecolor{magh}{RGB}{197,0,90}
\usepackage{cmbright}
\usepackage[OT1]{fontenc}

\usepackage[T1]{fontenc} % Schriftart
\usepackage[lining,light]{InriaSans}

\let\oldnormalfont\normalfont
\def\normalfont{\oldnormalfont\mdseries}

\usepackage[pagewise,modulo]{lineno}
\begin{document}

\title{Interplay of protection and damage through intermolecular processes in the decay of electronic core holes in microsolvated organic molecules}

\author{Dana~Blo\ss}
\email{dana.bloss@uni-kassel.de}
	\affiliation{Institute of Physics and Center for Interdisciplinary Nanostructure Science and Technology (CINSaT), University of Kassel, Heinrich-Plett-Straße 40, 34132 Kassel, Germany}
	
\author{Nikolai~V.~Kryzhevoi}
	\affiliation{Theoretische Chemie, Institut f\"ur Physikalische Chemie, Universit\"at Heidelberg, Im Neuenheimer Feld 229, 69120 Heidelberg, Germany}
 
\author{Jonas~Maurmann}
	\affiliation{Theoretische Chemie, Institut f\"ur Physikalische Chemie, Universit\"at Heidelberg, Im Neuenheimer Feld 229, 69120 Heidelberg, Germany}

\author{Philipp~Schmidt}
	\affiliation{European XFEL, Holzkoppel 4, 22869 Schenefeld, Germany}
 
\author{André~Knie}
	\affiliation{Institute of Physics and Center for Interdisciplinary Nanostructure Science and Technology (CINSaT), University of Kassel, Heinrich-Plett-Straße 40, 34132 Kassel, Germany}

\author{Johannes~H.~Viehmann}
	\affiliation{Institute of Physics and Center for Interdisciplinary Nanostructure Science and Technology (CINSaT), University of Kassel, Heinrich-Plett-Straße 40, 34132 Kassel, Germany}
 
\author{Sascha~Deinert}
	\affiliation{Deutsches Elektronen-Synchrotron DESY, Notkestraße 85, 22607 Hamburg, Germany}

\author{Gregor~Hartmann}
	\affiliation{Helmholtz-Zentrum Berlin (HZB), Albert-Einstein-Straße 15, D-12489 Berlin, Germany} 
 
\author{Catmarna~Küstner-Wetekam}
	\affiliation{Institute of Physics and Center for Interdisciplinary Nanostructure Science and Technology (CINSaT), University of Kassel, Heinrich-Plett-Straße 40, 34132 Kassel, Germany}

\author{Florian~Trinter}
	\affiliation{Fritz-Haber-Institut der Max-Planck-Gesellschaft, Faradayweg 4-6, 14195 Berlin, Germany}
	\affiliation{Institut f\"ur Kernphysik, Goethe-Universit\"at Frankfurt, Max-von-Laue-Straße\,1, 60438 Frankfurt am Main, Germany}
 
\author{Lorenz~S.~Cederbaum}
	\affiliation{Theoretische Chemie, Institut f\"ur Physikalische Chemie, Universit\"at Heidelberg, Im Neuenheimer Feld 229, 69120 Heidelberg, Germany}
	
\author{Arno~Ehresmann}
	\affiliation{Institute of Physics and Center for Interdisciplinary Nanostructure Science and Technology (CINSaT), University of Kassel, Heinrich-Plett-Straße 40, 34132 Kassel, Germany}

\author{Alexander~I.~Kuleff}
	\affiliation{Theoretische Chemie, Institut f\"ur Physikalische Chemie, Universit\"at Heidelberg, Im Neuenheimer Feld 229, 69120 Heidelberg, Germany}

\author{Andreas~Hans}
\email{hans@physik.uni-kassel.de}
	\affiliation{Institute of Physics and Center for Interdisciplinary Nanostructure Science and Technology (CINSaT), University of Kassel, Heinrich-Plett-Straße 40, 34132 Kassel, Germany}

\date{\today}

\begin{abstract}
Soft X-ray irradiation of molecules causes electronic core-level vacancies through photoelectronemission. In light elements, such as C, N, or O, which are abundant in the biosphere, these vacancies predominantly decay by Auger emission, leading inevitably to dissociative multiply charged states. It was recently demonstrated that an environment can prevent fragmentation of core-level-ionized small organic molecules through immediate non-local decay of the core hole, dissipating charge and energy to the environment. Here, we present an extended photoelectron-photoion-photoion coincidence (PEPIPICO) study of the biorelevant pyrimidine molecule embedded in a water cluster. It is observed and supported by theoretical calculations that the supposed protective effect of the environment is partially reversed if the vacancy is originally located at a water molecule. In this scenario, intermolecular energy or charge transfer from the core-ionized water environment to the pyrimidine molecule leads to ionization of the latter, however, presumably in non-dissociative cationic states. Our results contribute to a more comprehensive understanding of the complex interplay of protective and harmful effects of an environment in the photochemistry of microsolvated molecules exposed to X-rays.

\end{abstract}

\maketitle

\begin{figure}[h!]
\includegraphics[width=.48\textwidth]{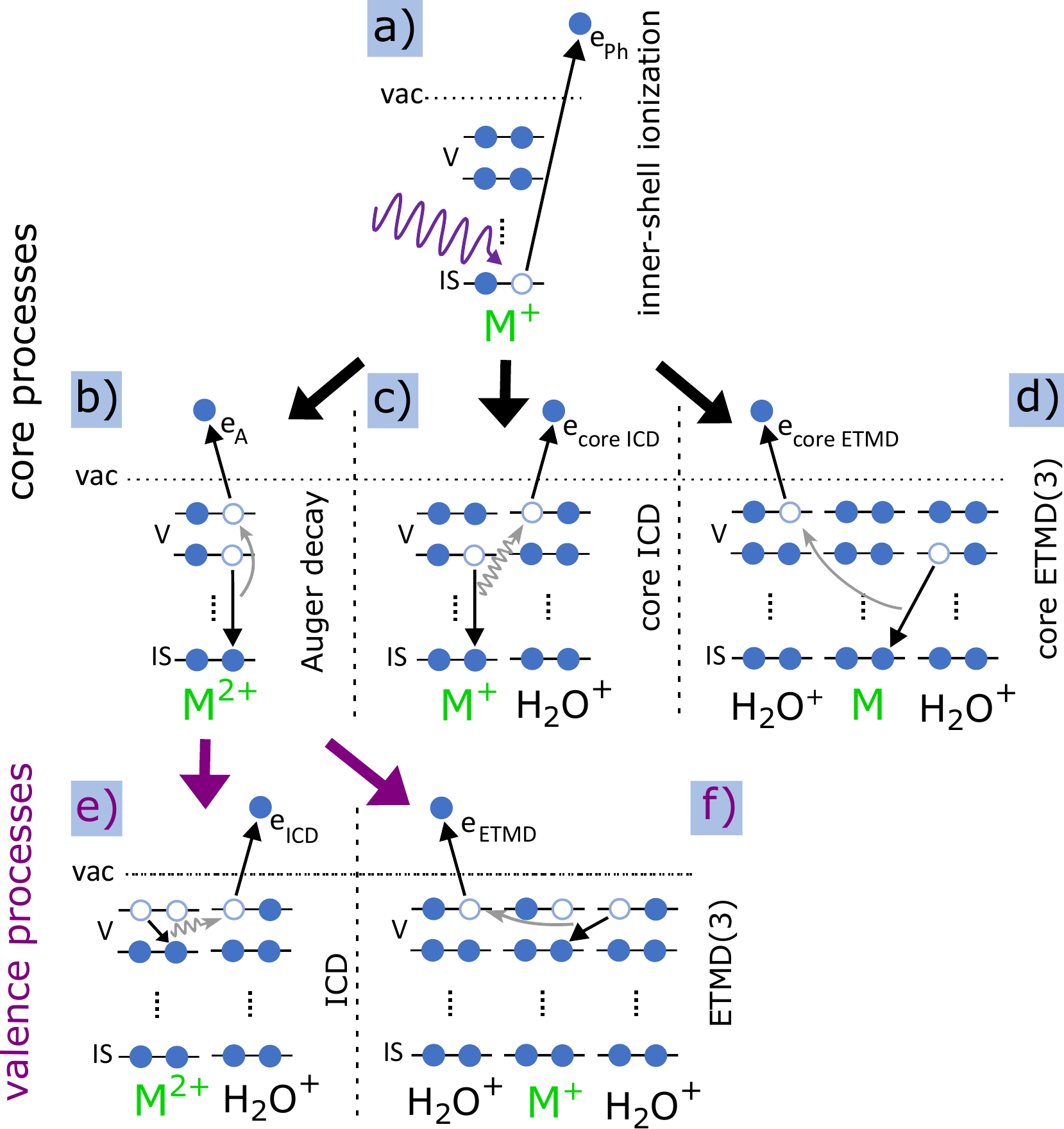}
  \caption{Sketched local and intermolecular processes after inner-shell photoionization of pyrimidine (M) embedded in a water cluster, grouped in mechanisms involving the core [a)-d)] or valence states [e) and f)]. Process a) is the inner-shell (IS) photoionization of a molecule M emitting a photoelectron e$_{\text{Ph}}$. Panel\,b) displays the subsequent Auger decay where a valence (V) electron fills the core vacancy and another valence electron from the same molecule is emitted (Auger electron e$_{\text{A}}$). Process c) shows the core-level ICD, transferring the released energy to a neighboring water molecule and ionizing a valence electron (e$_{\text{core ICD}}$). Process d sketches core-level ETMD(3). Here, a valence electron from neighboring water fills the inner-shell vacancy of the initially ionized molecule and a valance electron (e$_{\text{core ETMD}}$) from another water is released. ICD and ETMD can additionally happen after a preceding Auger decay [process e) and f)].}
  \label{fig:Ubergange_ETMD_ICD}
\end{figure}

X-ray-induced electron and nuclear dynamics are key topics in research on molecular radiation damage. In particular, understanding the fundamentals of radiation damage to DNA or its constituents is of utmost importance in medicine and biology \cite{Alizadeh2015, Dong2021,Sauer2022,Alizadeh2013, Bald2012a, Caron2009, Orlando2008, Xu2018}. The experimental investigation of x-ray-induced processes of DNA building blocks on a molecular level, however,  is challenging, if carried out in its natural environment. This environment has a decisive impact on the damage introduced to a biomolecule after interaction. The formation of secondary low-energy electrons \cite{Alizadeh2015, Sanche2009,Boudaiffa2000,Huels2003, Bloß2024, Gopakumar2023}, radicals \cite{Alizadeh2015, Sanche2009}, or ions \cite{Maclot2011,Lopez-Tarifa2013,Markush2016, Castrovilli2017, Nomura2017} typically are harmful for DNA, whereas the suppression of dissociation processes due to a liquid environment by steric hindrances or neutralisation of charges may protect DNA \cite{Hans2021, Johny2024}. In essence, therefore, the net effect of an environment for radiation damage is still not fully clear. \\
Inner-shell ionization of an isolated pyrimidine molecule (C$_{4}$H$_{4}$N$_{2}$), for example, initiates Auger decay, leaving the molecule in a doubly positively charged and therefore dissociative state. The result is fragmentation into two singly charged fragments \cite{Itala2011, Lin2015, Chiang2018,Bolognesi2012}. In  contrast, for inner-shell-ionized pyrimidine embedded in a water environment, intermolecular charge- and energy-transfer channels with neighbouring molecules  open up which compete with the local Auger decay, preventing the formation of doubly charged states in the molecule, and, therefore, its dissociation due to Coulomb repulsion \cite{Hans2021}. While experimentally the effect has not yet been quantified, theory predicts that already solvation by four water molecules is enough for the intermolecular channels to outpace Auger decay. Importantly, intermolecular decay of the inner-shell vacancy leaves the biomolecule intact in a neutral or singly charged state \cite{Cederbaum1997, stoychev2011}.
The intermolecular decay processes of electronic core-level vacancies considered here are energy- or charge-transfer mechanisms like core-level interatomic/intermolecular Coulombic decay (core-level ICD) \cite{Pokapanich2011, Hans2020, Slavicek2014} and core-level electron-transfer-mediated decay (core-level ETMD) \cite{Slavicek2014}.\\
Figure\,\ref{fig:Ubergange_ETMD_ICD} illustrates the relevant processes, starting with the inner-shell ionization [panel a)], which leaves the molecule singly charged and highly excited  (M$^{+}$).  Subsequently, local Auger decay [Fig.~\ref{fig:Ubergange_ETMD_ICD}\,b)] may take place, where the core hole is filled by a valence electron and another valence electron is ejected, leaving the originally singly charged ion in a doubly charged state (M$^{2+}$).  Competing with Auger decay, the released energy from the filling of the core vacancy can ionize a valence electron of a neighboring molecule via core-level ICD [Fig.~\ref{fig:Ubergange_ETMD_ICD}\,c)]. Core-level ICD produces a water cation and leaves both molecules singly charged. Alternatively, core-level ETMD [Fig.~\ref{fig:Ubergange_ETMD_ICD}\,d)] can take place. Two variants of ETMD are possible: ETMD(2) or ETMD(3), depending on the number of involved molecules \cite{Zobeley2001}. Figure\,\ref{fig:Ubergange_ETMD_ICD} only sketches ETMD(3), typically dominating over ETMD(2). In the former, one water-neighbor valence electron fills the core hole in the initially ionized molecule and the released energy is transferred to yet another water emitting one of its valence electrons. Consequently, ETMD(3) produces two water cations and a neutral, initial ionized molecule. In ETMD(2), the electron filling the vacancy and the electron being emitted originate from the same water molecule, leaving it in a doubly charged state.
% The charge of the initially ionized molecule is reduced in both cases by one unit, i.e., in the present example it fully neutralizes. 
Note that while core-level ICD has been observed in various systems, to our knowledge no experimental signature of core-level ETMD has been reported yet. In addition to the direct decay of the inner-shell hole, ICD and ETMD can take place also from Auger final states if the internal excess energy permits [Fig.~\ref{fig:Ubergange_ETMD_ICD}\,e) and f)] \cite{Gokhberg2021-163, Bloß2024, Gopakumar2023}. Here, instead of a core-level, a valence-level vacancy is filled. While the kinetic energy of electrons emitted in core-level ICD or ETMD are comparable to Auger electron energies, ICD or ETMD electrons from Auger final states have considerably less energy, typically between zero to a few tens of eV \cite{stoychev2011,Skitnevskaya2023}.\\

While in the case of inner-shell ionization of pyrimidine the intermolecular processes “protect” the molecule from fragmentation by energy and charge transfer to the water surrounding \cite{Hans2021}, it is now an intriguing question what happens if the X-ray photon targets the water environment instead. Will intermolecular processes in turn lead to ionization and fragmentation of the pyrimidine? To address these questions, we core-ionized the water in a heterogeneous pyrimidine-water cluster, recorded the resulting photoelectron-photoion-photoion coincidence (PEPIPICO) spectra and combined the measurement  with calculated electron spectra. 

\begin{figure}
\includegraphics[width=.45\textwidth]{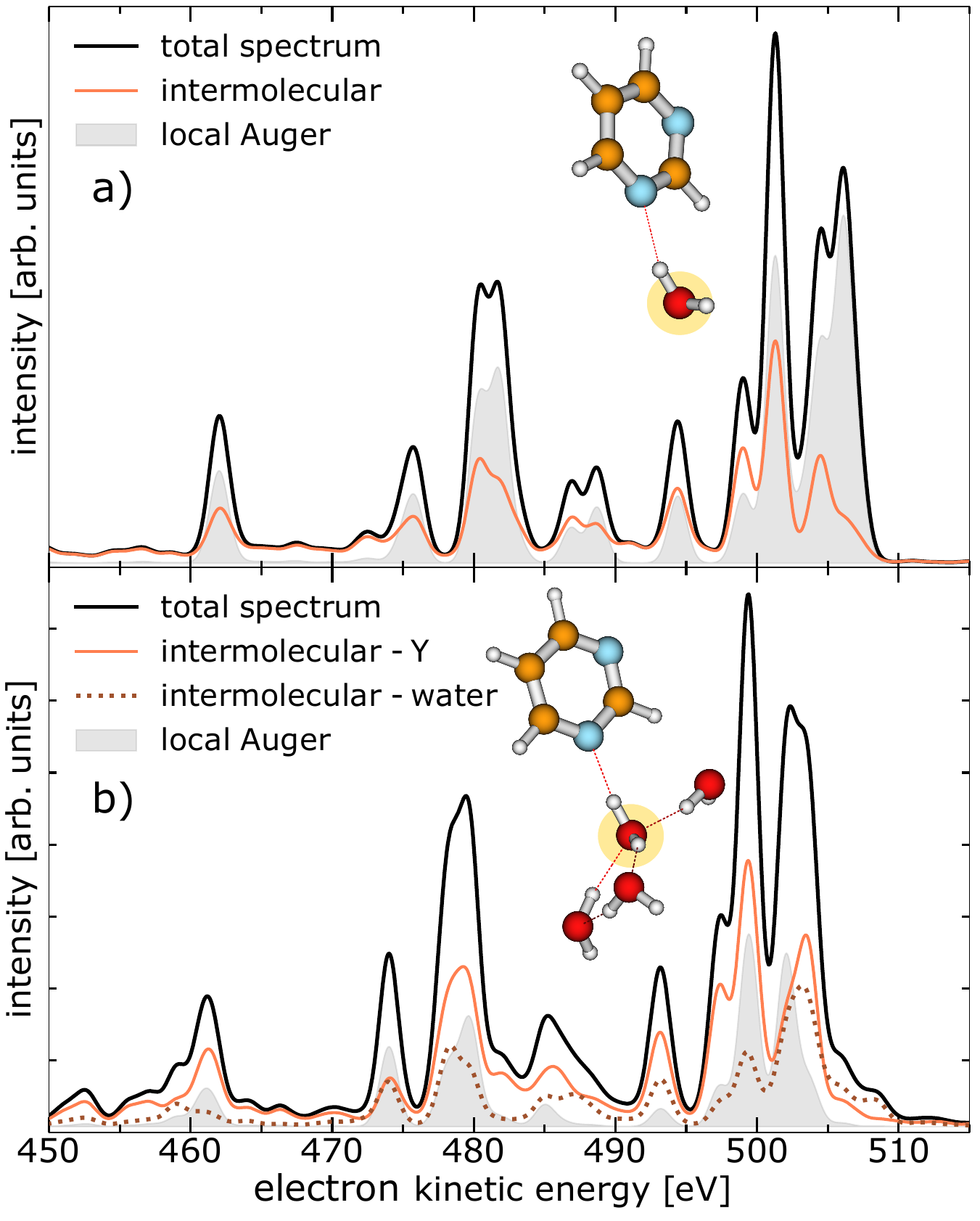}
  \caption{Theoretical calculation of the kinetic-energy spectrum of a pyrimidine-H$_{2}$O dimer in panel\,a) and a pyrimidine-(H$_{2}$O)$_{4}$ cluster in panel\,b) after O\,1s photoionization, broken down by the nature of the final states. The calculated equilibrium geometry of the clusters is shown and the targeted O atom is yellow marked. The total electron emissions are shown (black trace) with the contributions of the local Auger decay (filled gray trace) and intermolecular processes involving the pyrimidine,  abbreviated as Y, (orange solid line). The intermolecular processes only between water molecules are presented as brown dotted line in panel\,b).}
  \label{fig:Theorie}
\end{figure}

Similar to Ref.\,\cite{Hans2021}, we performed calculations on the strengths of the different local and non-local decay channels contributing to the decay of the O\,1s vacancy in pyrimidine-(H$_{2}$O)$_{n}$ complexes. The simulated electron emission spectra are displayed in Fig.\,\ref{fig:Theorie}, for one water molecule in the vicinity of the pyrimidine in panel\,a), and for four water molecules in panel\,b). The equilibrium geometries of the two pyrimidine-water complexes are illustrated in the insets, with the core-ionized O atom marked in yellow. The kinetic energy spectrum of all expected electrons (black solid lines)  is shown with the contributions of the  local Auger decay (filled gray traces), the intermolecular processes involving pyrimidine (orange solid lines), and the intermolecular processes only including  water molecules [brown dotted line in panel\,b)]. As is evident from Fig.\,\ref{fig:Theorie}, the intermolecular processes exhibit already a considerable share with only one water present a) and outpace the local processes clearly in the presence of four water molecules b). In the Supplementary Information, the calculated electron spectra for pyrimidine in the vicinity of one to four water molecules are shown in more detail, including a breakdown of the intermolecular processes in their contributions of core-level ICD, core-level ETMD(2), and core-level ETMD(3).

\begin{figure}[h!]
\includegraphics[width=.45\textwidth]{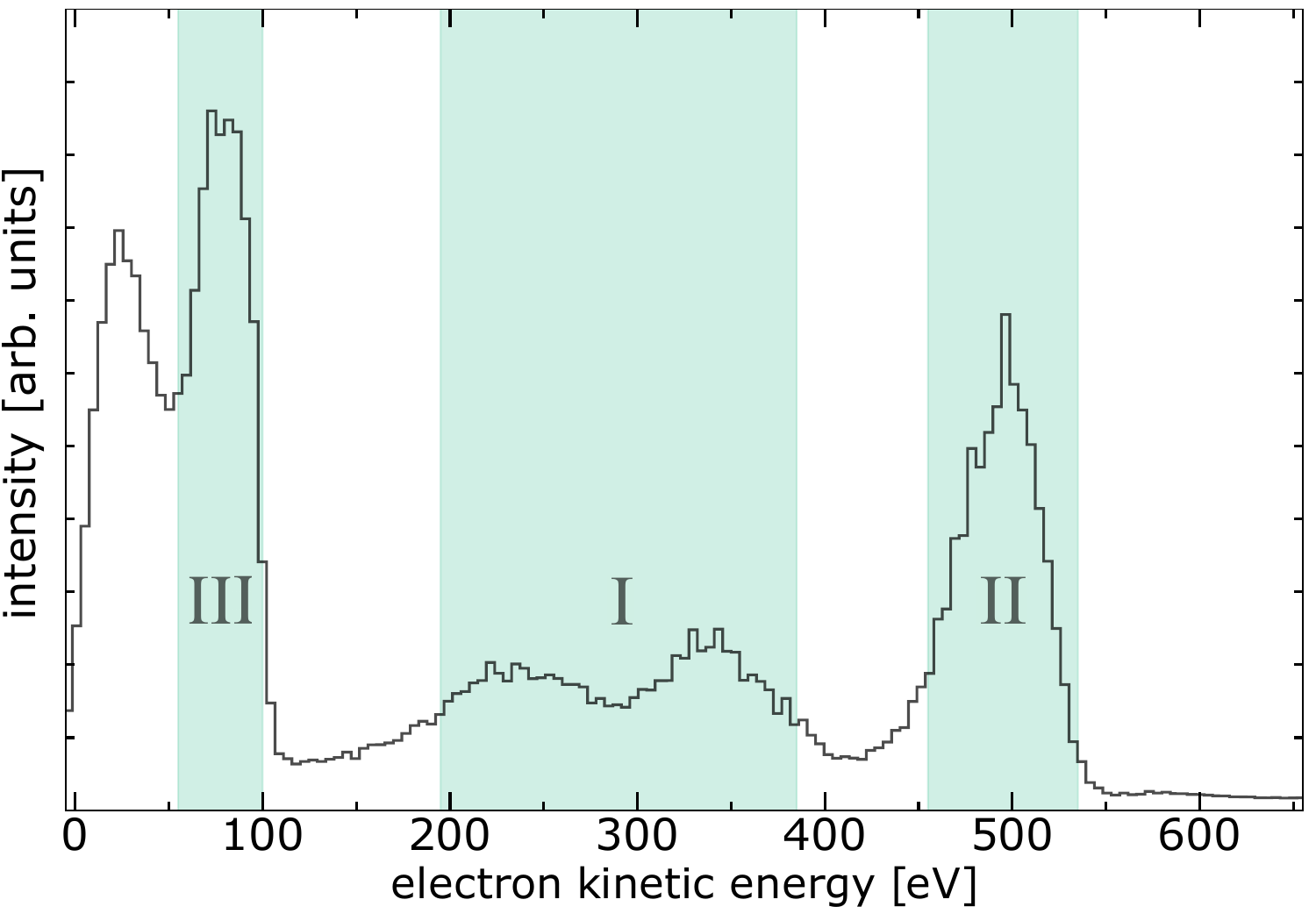}
  \caption{Electron spectrum from PEPIPICO of a mixed pyrimidine-water target at an exciting-photon energy of 620\,eV. Three regions are indicated with electrons originating from different processes used as filter conditions for the ion-ion coincidence maps in Figs.\,\ref{fig:Coin-Maps}\,a)-c).}
  \label{fig:Elektronenspektrum}
\end{figure}

Experimentally, we measured PEPIPICO spectra from pyrimidine-water clusters at an exciting-photon energy of 620\,eV. The corresponding electron spectrum is shown in Fig.\,\ref{fig:Elektronenspektrum}. The relatively poor resolution of the spectrum is a result of the applied extraction voltages required for the ion detection (see Methods section and Ref.\,\cite{Hans2021} for details). Nevertheless, it is sufficient to distinguish certain features: the O\,1s photoelectrons (region {\fontfamily{ptm}\selectfont III} in Fig.\,\ref{fig:Elektronenspektrum}, 55-100\,eV), electrons related to emission from pyrimidine, both inner-shell photoelectrons and secondary electrons (region {\fontfamily{ptm}\selectfont I}, 195-385\,eV), and electrons ejected in the decay of water core-level vacancies (region {\fontfamily{ptm}\selectfont II}, 455-535\,eV). 
In addition to these three main features, one very weak maximum close to 600\,eV contains all photoemitted valence electrons. Finally, another peak around 20\,eV can be identified, which is mainly formed by high-energy electrons (e.g., photoelectrons or Auger electrons) that have lost a substantial amount of their energy through inelastic scattering, by electrons created from electron-impact ionization, and by electrons emitted through valence-level ICD or ETMD. Note that other weak processes with non-discrete electron spectrum like double Auger decay may lie below the main features. For region {\fontfamily{ptm}\selectfont III} (55-100\,eV) associated to the O\,1s photoelectron peak, with a nominal kinetic energy at about 82\,eV \cite{Bjorneholm1999,Ohrwall2005,Tenorio2018},  the signal of slow electrons extends significantly into this region. In region {\fontfamily{ptm}\selectfont I} (195-385\,eV), we expect photoelectrons and Auger electrons emitted after the C\,1s (binding energies of 291.09\,eV, 292.08\,eV, and 292.48\,eV \cite{Bolognesi2010a}) and N\,1s (binding energy of 405.23\,eV \cite{Bolognesi2010}) ionization of pyrimidine as well as electrons related to non-local processes initiated by these inner-shell ionizations \cite{Hans2021}. Electrons in  region {\fontfamily{ptm}\selectfont II} (455-535\,eV) result from Auger and non-local decays after core ionization of water, either of free, gaseous water contained in the jet, or bound in homogeneous (H$_{2}$O)$_{n}$ or heterogeneous pyrimidine-(H$_{2}$O)$_{n}$ clusters.

\begin{figure*}[!]
\includegraphics[width=\textwidth]{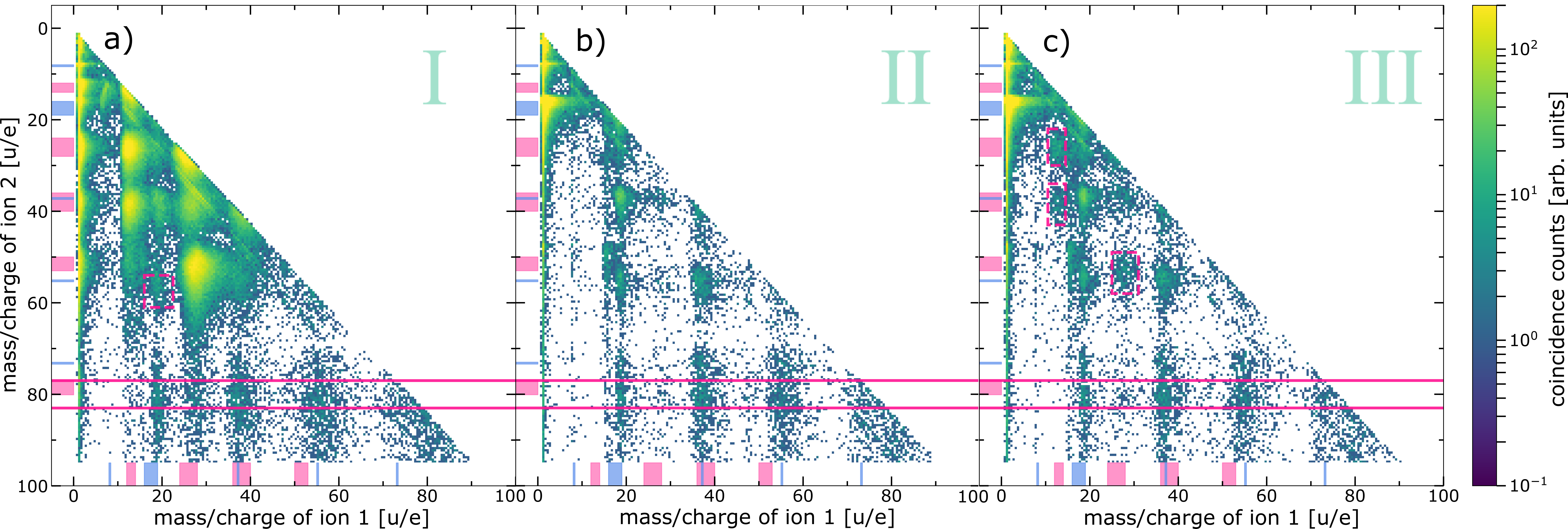}
  \caption{Ion-ion coincidence maps recorded at an exciting-photon energy of 620\,eV coincidently measured with electrons of different kinetic energy ranges. a) Ion-ion pairs in coincidence with electrons resulting from pyrimidine core-level processes (region\,{\fontfamily{ptm}\selectfont I} in Fig.\,\ref{fig:Elektronenspektrum}), b) with water core-level processes (region\,{\fontfamily{ptm}\selectfont II} in Fig.\,\ref{fig:Elektronenspektrum}, excluding the O\,1s photoelectrons), and c) with the O\,1s photoelectrons (region\,{\fontfamily{ptm}\selectfont III} in Fig.\,\ref{fig:Elektronenspektrum}). Note the logarithmic color scale. The ranges of expected fragments originating from pyrimidine (pink) and water monomers or clusters (blue) according to Table\,\ref{tab:Y-H2O Fragemte} are indicated with respective markers at both axes. The horizontal pink solid lines indicate the region of coincidences of an intact pyrimidine parent ion ($m/q_{ion2}$ = 80 ± 3\,u/e)  and another ion. Some additional features are highlighted with pink dashed boxes, for details see text.}
  \label{fig:Coin-Maps}
\end{figure*}

The mass-to-charge ($m/q$) ratios of the main fragment ions resulting from inner-shell ionization of water or pyrimidine are listed in Table\,\ref{tab:Y-H2O Fragemte}. As evident from the table, expected features from some of the ion fragments overlap or are close in their $m/q$ ratio making some assignments challenging. This is not the case, however, for the pyrimidine parent ion with a $m/q$ of 80\,u/e.  The corresponding maps of ion-ion pairs selected for coincidences with electrons of the three indicated regions are shown in Figs.\,\ref{fig:Coin-Maps}\,a)-c).

\begin{table}
\centering
\begin{tabular}{l|l}
       \multicolumn{1}{c|}{Mass-to-charge}  &\multicolumn{1}{c}{\multirow{2}{*}{Corresponding ions}}  \\
       \multicolumn{1}{c|}{ratio [u/e]}  &  \\
        \noalign{\hrule height .8pt}
       1 &       H$^{+}$ \\
       \rowcolor[gray]{.9} 8 &       O$^{2+}$ \\
       12-14 & C$^{+}$ / CH$^{+}$ /  N$^{+}$ \cite{Lin2015} \\
       \rowcolor[gray]{.9} 16-19  & O$^{+}$ / OH$^{+}$ / H$_{2}$O$^{+}$ / H$_{2}$OH$^{+}$ \\
       24-28 & C$_{2}$$^{+}$ / C$_{2}$H$^{+}$ / C$_{2}$H$_{2}$$^{+}$ / CN$^{+}$ / CHN$^{+}$ / CH$_{2}$N$^{+}$  \cite{Lin2015}\\
       \rowcolor[gray]{.9} 37 & (H$_{2}$O)$_{2}$H$^{+}$ \\
       
      \multirow{2}{*}{36-40}  & \multicolumn{1}{l}{C$_{3}$$^{+}$ / C$_{3}$H$^{+}$ / C$_{3}$H$_2^{+}$ / C$_{3}$H$_{3}$$^{+} /  $C$_{2}$N$^{+}$  / C$_{2}$HN$^{+}$ /}\\
       & \multicolumn{1}{l}{ C$_{2}$H$_{2}$N$^{+}$ / CN$_{2}$$^{+}$   \cite{Lin2015}}\\
       
     \rowcolor[gray]{.9}  & \multicolumn{1}{l}{C$_{3}$N$^{+}$ / C$_{3}$HN$^{+}$ / C$_{3}$H$_{2}$N$^{+}$ / C$_{3}$H$_{3}$N$^{+}$ / C$_{2}$N$_{2}$$^{+}$  /}\\
        \rowcolor[gray]{.9} \multirow{-2}{*}{50-53}   & \multicolumn{1}{l}{C$_{2}$HN$_{2}$$^{+}$  \cite{Lin2015}}\\
       
       55 & (H$_{2}$O)$_{3}$H$^{+}$ \\
       
       \rowcolor[gray]{.9} 80 & C$_{4}$H$_{4}$N$_{2}$$^{+}$ \,\,(\textcolor{magh}{pyrimidine parent ion})
\end{tabular}
\caption{Mass-to-charge ($m/q$) ratio of the main fragment ions expected from inner-shell ionization of a pyrimidine-water mixture. The $m/q$ ratio of the pyrimidine fragments was taken from\,\cite{Lin2015}. Many of the pyrimidine fragments and water-cluster ions overlap in their $m/q$ ratios. The intact pyrimidine ion can be found at a $m/q = 80\,\text{u/e}$.}
\label{tab:Y-H2O Fragemte}
\end{table}

Figure\,\ref{fig:Coin-Maps}\,a) exhibits all events of two ions in coincidence with electrons of region {\fontfamily{ptm}\selectfont I} of Fig.\,\ref{fig:Elektronenspektrum}, emitted after inner-shell ionization of the C or N atoms of the pyrimidine. A similar ion-ion map can be found in Ref.\,\cite{Hans2021} recorded below the N\,1s and O\,1s edges. Both maps mainly show features originating from the fragmentation of pyrimidine into two cationic fragments ($m/q = 12-14\,\text{u/e}, 24-28\,\text{u/e}, 36-40\,\text{u/e}, \text{and~
} 50-53\,\text{u/e}$). These breakup channels agree well with previously reported fragment spectra of inner-shell-ionized gas-phase pyrimidine \cite{Lin2015, Bolognesi2012, Bolognesi2010a}. Besides these main features, some additional observations are noteworthy: (i)\,one pyrimidine fragment in coincidence with a water-cluster ion ($m/q $ = 16-19\,u/e, 37\,u/e, and 55\,u/e), (ii)\,one intact pyrimidine molecule ($m/q$ of 80\,u/e) with water-cluster ions, and (iii)\,one intact pyrimidine molecule with pyrimidine fragments. These observations have been discussed previously and case\,(ii) was interpreted as a protective effect of the water surrounding the biomolecule due to the intermolecular core-level processes \cite{Hans2021}. Another weak but interesting feature is the coincident emission of an ion pair with ion\,1: $m/q = 16-19$\,u/e and ion\,2: $m/q = 55$\,u/e, both assigned to water-cluster ions and highlighted with a pink box in Fig.\,\ref{fig:Coin-Maps}\,a). This feature has not been discussed before and provides evidence for core-level ETMD(3): despite C\,1s or N\,1s ionization, no pyrimidine fragment is observed, but two water-cluster ions are formed instead. This interpretation is, however, tentative, because the signal is weak and the resolution is poor. An alternative way to produce two water cations is valence ETMD(3) subsequent to Auger decay, if the pyrimidine parent ion is missed in detection.

The ion-ion map of Fig.\,\ref{fig:Coin-Maps}\,b) contains ion pairs in coincidence with an electron emitted in a water O\,1s decay. In comparison to Fig.\,\ref{fig:Coin-Maps}\,a) the overall map differs significantly, mainly because of the absence of two pyrimidine fragment ions in coincidence. The main feature originates here from coincidences of two fragments of the water monomer. As expected, we also observe signal resulting from two water-cluster fragments. Interestingly, there is significant signal from an intact pyrimidine ion with water-cluster fragments (between pink solid lines). Therefore, we can conclude that an intermolecular energy or charge transfer has taken place initiated by an O\,1s ionization and ending with a singly ionized pyrimidine. Surprisingly, no clear signal of pyrimidine fragments in coincidence with any other ion can be found in this ion-ion map.

Figure\,\ref{fig:Coin-Maps}\,c) depicts the ion-ion map in coincidence with the O\,1s photoelectron. Intuitively, this map should contain the same features as the ion-ion map in coincidence with the water core-level decay processes, i.e., Fig.\,\ref{fig:Coin-Maps}\,b). By comparing both maps, we can identify three features present in Fig.\,\ref{fig:Coin-Maps}\,c) that are absent in Fig.\,\ref{fig:Coin-Maps}\,b), highlighted by pink dashed boxes. All of them can be assigned to one or two pyrimidine fragments. Taking into account the electron spectrum shown in Fig.\,\ref{fig:Elektronenspektrum}, these pyrimidine fragments may result from coincidences with slow electrons (originating from pyrimidine or water) instead of coincidences with the O\,1s photoelectrons.
Owing to experimental challenges, the resolution in both electron and ion spectra is relatively low. The different decay channels can thus not be compared quantitatively. However, the presence of pyrimidine parent ions but absence of pyrimidine fragment ions in Fig.\,\ref{fig:Coin-Maps}\,b) allows some further conclusions. First, ETMD(2) of water O\,1s vacancies with pyrimidine seems to be unlikely, since it would  lead to doubly charged dissociative pyrimidine. Second, the observation of intact pyrimidine parent ions but no fragments is an evidence for core-level ICD or core-level ETMD(3), or ICD/ETMD(3) from Auger final states. In all cases, it seems that predominantly the four outermost valence orbitals of pyrimidine are ionized. These valence-ionized states are known to be stable, while deeper valence ionization is dissociative \cite{Plekan2008}. For an improved disentanglement and unambiguous (quantitative) assignment of local or non-local processes, a higher resolution in the electron spectrum of the PEPIPICO is crucial. Especially in combination with theory, the investigation and understanding of these decays can enhance our knowledge of the mechanisms of radiation damage on a molecular level. The distinction of non-local mechanisms, taking place directly in the decay of the core-level vacancy, and those from excited valence states subsequent to Auger decay seems feasible with high-resolution electron spectra. In this regard, high-resolution studies tackling this topic would be worthwhile to conduct.

Summarizing, we measured PEPIPICO of core-ionized pyrimidine-water clusters. Taking advantage of the coincidence technique, we were able to distinguish clusters with core holes located on the pyrimidine or the water molecule and to analyze the resulting ions created in the decay. A previously described energy transfer from ionized pyrimidine to neighboring water via non-local core-hole decay could be confirmed. Additionally, we identified the opposite scenario. Theoretically and experimentally, we observe water core-level ionization and subsequent immediate or cascade non-local decay involving neighboring pyrimidine or other water molecules. Interestingly, the pyrimidine molecules seem to end solely in singly charged non-dissociative states, as the absences of pyrimidine fragments are suggesting. Additionally, we found first experimental indications for a core-level ETMD process after initial C\,1s/N\,1s ionization of the pyrimidine molecule in a cluster.

\section{Methods}
The experiment was performed at the P04 soft X-ray beamline of PETRA\,III (DESY, Hamburg, Germany). A photoelectron-photoion-photoion coincidence (PEPIPICO) setup was used, following the scheme of Ref.\,\cite{Eland2006}. The setup consists of a magnetic-bottle time-of-flight electron spectrometer, equipped with a ring-shaped permanent magnet, an approximately 910\,mm long drift tube, where several retardation voltages can be applied, and a microchannel-plate detector. Opposite to the electron drift tube, a 23\,mm long ion time-of-flight spectrometer is mounted. The ring-shaped permanent magnet is part of the ion spectrometer, extracting the ions towards their drift tube. The synchrotron was operated in the 40-bunch mode with 192\,ns time spacing between two consecutive light pulses. The light beam was crossed orthogonally with a pyrimidine-water cluster jet in the horizontal plane and the ion and electron spectrometers were operated vertically. The cluster jet was produced by the supersonic coexpansion of a mixture through a conical 80\,µm nozzle with an opening angle of 30°. The mixture consists of pyrimidine-water vapor resulting from evaporation of a liquid mixture, heated to 80°\,C and containing 94\,\% water and 6\,\% pyrimidine. The cluster jet consists of water and pyrimidine monomers, pure water clusters, water-pyrimidine clusters, and in very small quantities pure pyrimidine clusters. The expansion chamber was separated from the interaction chamber through a skimmer with a 0.7\,mm opening. For a detailed description of the experiment as well as data acquisition and treatment see Ref.\,\cite{Hans2021}.\\
The spectra of the electrons emitted by the decay of the O\,1s vacancy were computed following the well-established methodology, used also in our previous work, Ref.\,\cite{Hans2021}. Here, we briefly outline the main ingredients. The equilibrium geometries of the studied pyrimidine-water complexes were obtained using the second-order Møller-Plesset (MP2) method with the cc-pVTZ basis sets. The dicationic states populated by the decay of core-ionized water molecules in the pyrimidine-water complex were obtained using the statistical method for computing Auger spectra proposed in Ref.\,\cite{Tarantelli1991}. The dicationic states of the system were computed with the help of the second-order algebraic diagrammatic construction scheme [ADC(2)] for the calculation of the poles and residues of the particle-particle propagator \cite{Schirmer1984,Tarantelli1985} using the cc-pVDZ basis sets. Depending on the localization of the two final holes, each state in the dicationic spectrum was then decomposed into different contributions attributed to Auger (two holes on the initially ionized water molecule), ICD (one hole on the initially ionized water molecule and one hole on the pyrimidine or another water molecule), ETMD(2) (two holes on the pyrimidine or another water molecule), and ETMD(3) (two holes distributed on two molecules different from the initially ionized water). As the number of final dicationic states of the pyrimidine-water complex is enormous, computing the individual decay rates is out of reach. To estimate the corresponding decay rates, we thus used the contribution in each transition moment of the configuration with two holes in the oxygen atom bearing the initial vacancy. Finally, the kinetic energies of the electrons emitted in the O\,1s decay were obtained by subtracting the populated dicationic states from the energy of the O\,1s core-hole state. The latter was computed by the means of the $\Delta$SCF method using the cc-pVDZ basis sets. To account for the vibrational broadening and the experimental resolution, the resulting spectra were convoluted with a Gaussian function with FWHM of 1.5\,eV.

\bibliography{library, lib}
\bibliographystyle{naturemag}

\begin{acknowledgments}
\section{Acknowledgements}

We acknowledge DESY (Hamburg, Germany), a member of the Helmholtz Association HGF, for the provision of experimental facilities and allocation of beamtime for proposal I-20180199. We are grateful for the excellent support from the PETRA~III P04 beamline staff and to Miriam\,Gerstel, Clara\,M.\,Saak, Rebecca\,Schaf, Jens\,Buck and Stephan\,Klumpp for assistance during the beamtime. This work was supported by the German Federal Ministry of Education and Research (BMBF) through projects 05K22RK2 – GPhaseCC and 05K22RK1 – TRANSALP as well as SFB 1319 ELCH, funded by the Deutsche Forschungsgemeinschaft (DFG; project No. 328961117). We also acknowledge the scientific exchange and support of the Centre for Molecular Water Science (CMWS). F.\,T. acknowledges funding by the Deutsche Forschungsgemeinschaft (DFG, German Research Foundation) - Project 509471550, Emmy Noether Programme and acknowledges support by the MaxWater initiative of the Max-Planck-Gesellschaft. L.\,S.\,C. gratefully acknowledges financial support by the European Research Council (ERC) (Advanced Investigator Grant No. 692657). 
\end{acknowledgments}

\end{document}

\title{Supplementary Information - Interplay of protection and damage through intermolecular processes in the decay of electronic core holes in microsolvated organic molecules}

\author{Dana~Blo\ss}
\email{dana.bloss@uni-kassel.de}
	\affiliation{Institute of Physics and Center for Interdisciplinary Nanostructure Science and Technology (CINSaT), University of Kassel, Heinrich-Plett-Straße 40, 34132 Kassel, Germany}
	
\author{Nikolai~V.~Kryzhevoi}
	\affiliation{Theoretische Chemie, Institut f\"ur Physikalische Chemie, Universit\"at Heidelberg, Im Neuenheimer Feld 229, 69120 Heidelberg, Germany}
 
\author{Jonas~Maurmann}
	\affiliation{Theoretische Chemie, Institut f\"ur Physikalische Chemie, Universit\"at Heidelberg, Im Neuenheimer Feld 229, 69120 Heidelberg, Germany}

\author{Philipp~Schmidt}
	\affiliation{European XFEL, Holzkoppel 4, 22869 Schenefeld, Germany}
 
\author{André~Knie}
	\affiliation{Institute of Physics and Center for Interdisciplinary Nanostructure Science and Technology (CINSaT), University of Kassel, Heinrich-Plett-Straße 40, 34132 Kassel, Germany}

\author{Johannes~H.~Viehmann}
	\affiliation{Institute of Physics and Center for Interdisciplinary Nanostructure Science and Technology (CINSaT), University of Kassel, Heinrich-Plett-Straße 40, 34132 Kassel, Germany}
 
\author{Sascha~Deinert}
	\affiliation{Deutsches Elektronen-Synchrotron DESY, Notkestraße 85, 22607 Hamburg, Germany}

\author{Gregor~Hartmann}
	\affiliation{Helmholtz-Zentrum Berlin (HZB), Albert-Einstein-Straße 15, D-12489 Berlin, Germany}

\author{Catmarna~Küstner-Wetekam}
	\affiliation{Institute of Physics and Center for Interdisciplinary Nanostructure Science and Technology (CINSaT), University of Kassel, Heinrich-Plett-Straße 40, 34132 Kassel, Germany}

\author{Florian~Trinter}
	\affiliation{Fritz-Haber-Institut der Max-Planck-Gesellschaft, Faradayweg 4-6, 14195 Berlin, Germany}
	\affiliation{Institut f\"ur Kernphysik, Goethe-Universit\"at Frankfurt, Max-von-Laue-Straße\,1, 60438 Frankfurt am Main, Germany}
 
\author{Lorenz~S.~Cederbaum}
	\affiliation{Theoretische Chemie, Institut f\"ur Physikalische Chemie, Universit\"at Heidelberg, Im Neuenheimer Feld 229, 69120 Heidelberg, Germany}
	
\author{Arno~Ehresmann}
	\affiliation{Institute of Physics and Center for Interdisciplinary Nanostructure Science and Technology (CINSaT), University of Kassel, Heinrich-Plett-Straße 40, 34132 Kassel, Germany}

\author{Alexander~I.~Kuleff}
	\affiliation{Theoretische Chemie, Institut f\"ur Physikalische Chemie, Universit\"at Heidelberg, Im Neuenheimer Feld 229, 69120 Heidelberg, Germany}

\author{Andreas~Hans}
\email{hans@physik.uni-kassel.de}
	\affiliation{Institute of Physics and Center for Interdisciplinary Nanostructure Science and Technology (CINSaT), University of Kassel, Heinrich-Plett-Straße 40, 34132 Kassel, Germany}

\date{\today}

\maketitle

\section{Theoretical electron spectra }
The Figs.\,\ref{fig:P-1W}, \ref{fig:P-2W}, \ref{fig:P-3W}, and \ref{fig:P-4W} show the theoretical electron spectra after O\,1s ionization  of a pyrimidine-water cluster with one to four water molecules in detail.  The geometry of the clusters are shown as insets in the corresponding figures. 
In all figures, the black solid traces present the total electron spectra, the gray filled traces the  Auger spectra, the pink dotted traces the core-level ICD spectra, the blue dashed traces the core-level ETMD(2) spectra, and the green  dashed-dotted traces the core-level ETMD(3) spectra. The  spectra on the left side illustrate the intermolecular processes with holes at the pyrimidine in the final states and the spectra  on the right side the intermolecular processes with holes only at water molecules in the finals states. The different rows correspond to different ionized water molecules (if the cluster contains more than one water) and the ionized water is indicated by a black arrow. For a detailed description of the theoretical methods see the methods section of the main article. \\

\renewcommand{\thefigure}{S1}
\begin{figure}[h!]
\includegraphics[width=.5\textwidth]{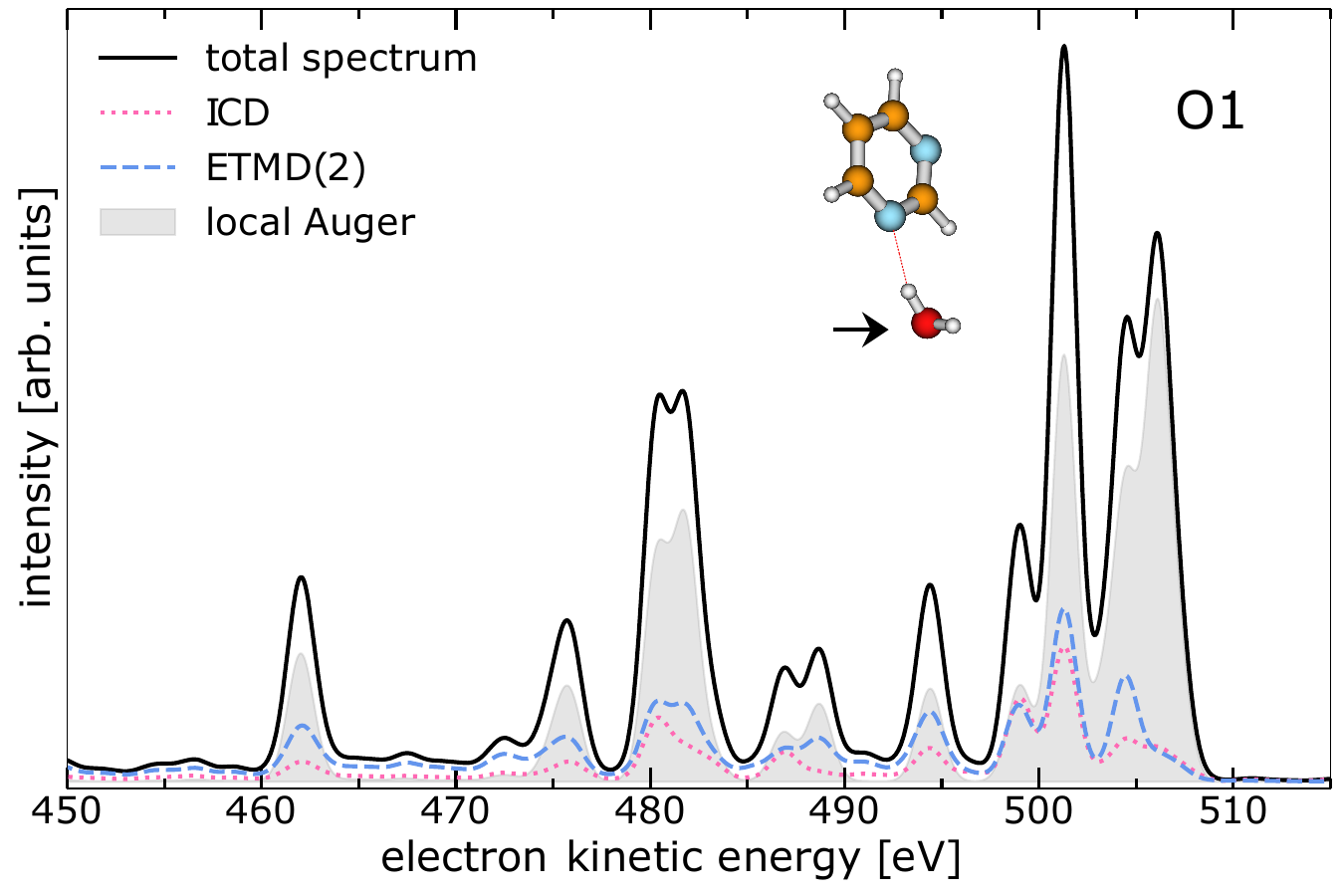}
  \caption{Theoretical spectra of the emitted electrons in the decay of an O\,1s vacancy in a pyrimidine-water dimer. The equilibrium geometry of this dimer is illustrated as an inset. The different local  (Auger decay - gray filled trace) and intermolecular  contributions (core-level ICD -  pink dotted trace and  core-level ETMD(2) -  blue dashed trace)  of the  total spectra (black solid trace) are presented. In the dimer, all non-local decay paths necessarily imply holes at the pyrimidine in the final state.}
  \label{fig:P-1W}
\end{figure}

\renewcommand{\thefigure}{S2}
\begin{figure}[h!]
\includegraphics[width=\textwidth]{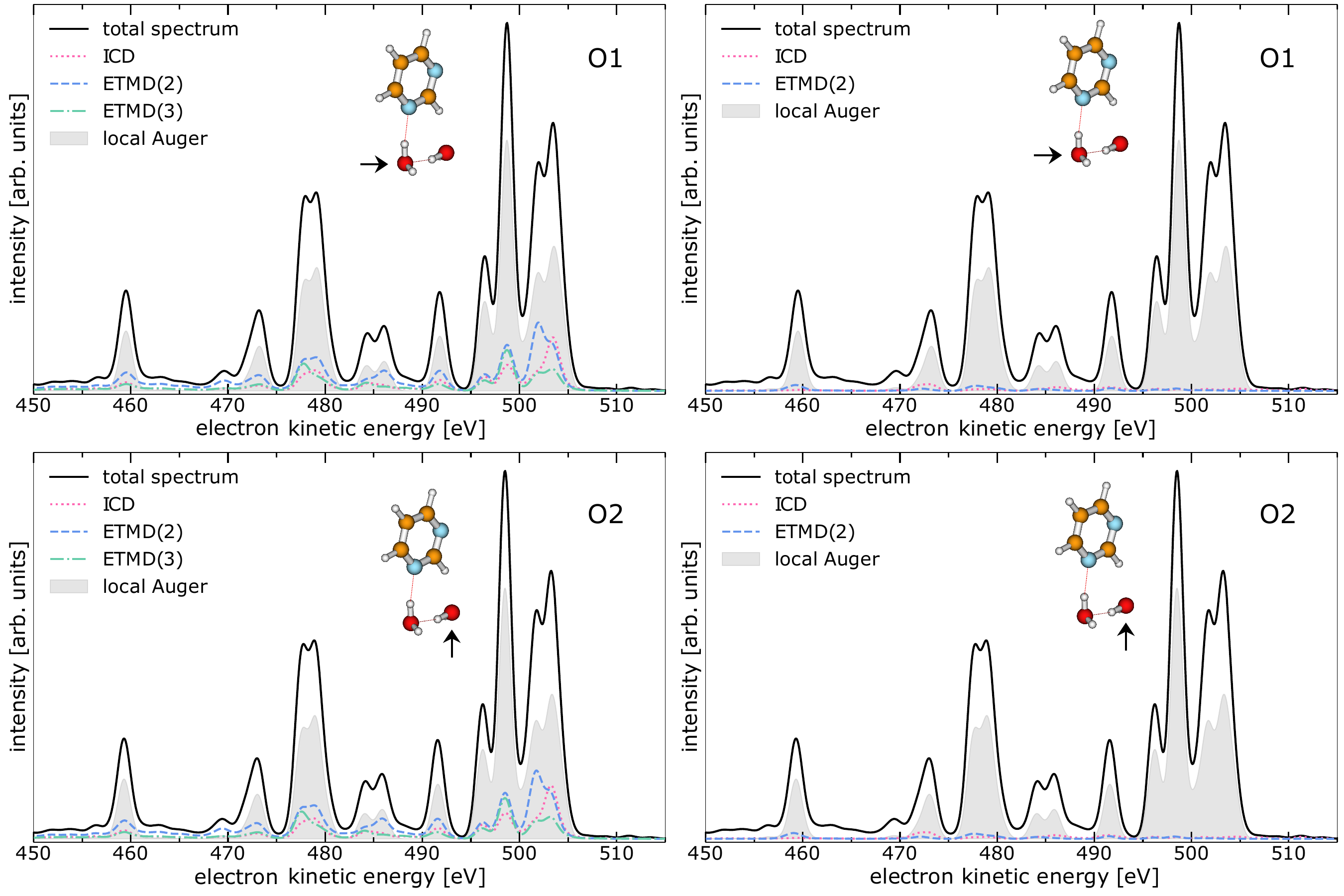}
  \caption{Theoretical spectra of the emitted electrons in the decay of an O\,1s vacancy in a pyrimidine-water cluster, consisting of one pyrimidine molecule and two water molecules. 
 The equilibrium geometry of the cluster is illustrated as  insets. The different local  (Auger decay - gray filled traces) and intermolecular contributions (core-level ICD -  pink dotted traces, core-level ETMD(2) -  blue dashed traces, and ETMD(3) - green dashed-dotted traces)  of the  total spectra (black solid  traces) are presented. The spectra on the left side show the intermolecular processes with holes at the pyrimidine in the final states and the spectra  on the right side the intermolecular processes with holes only at water molecules in the finals states. Note that identical total and Auger spectra are shown in all figures for comparison. The different rows depict  spectra resulting from the ionization of different waters (marked with black arrow). }
  \label{fig:P-2W}
\end{figure}

\renewcommand{\thefigure}{S3}
\begin{figure}[h!]
\includegraphics[width=\textwidth]{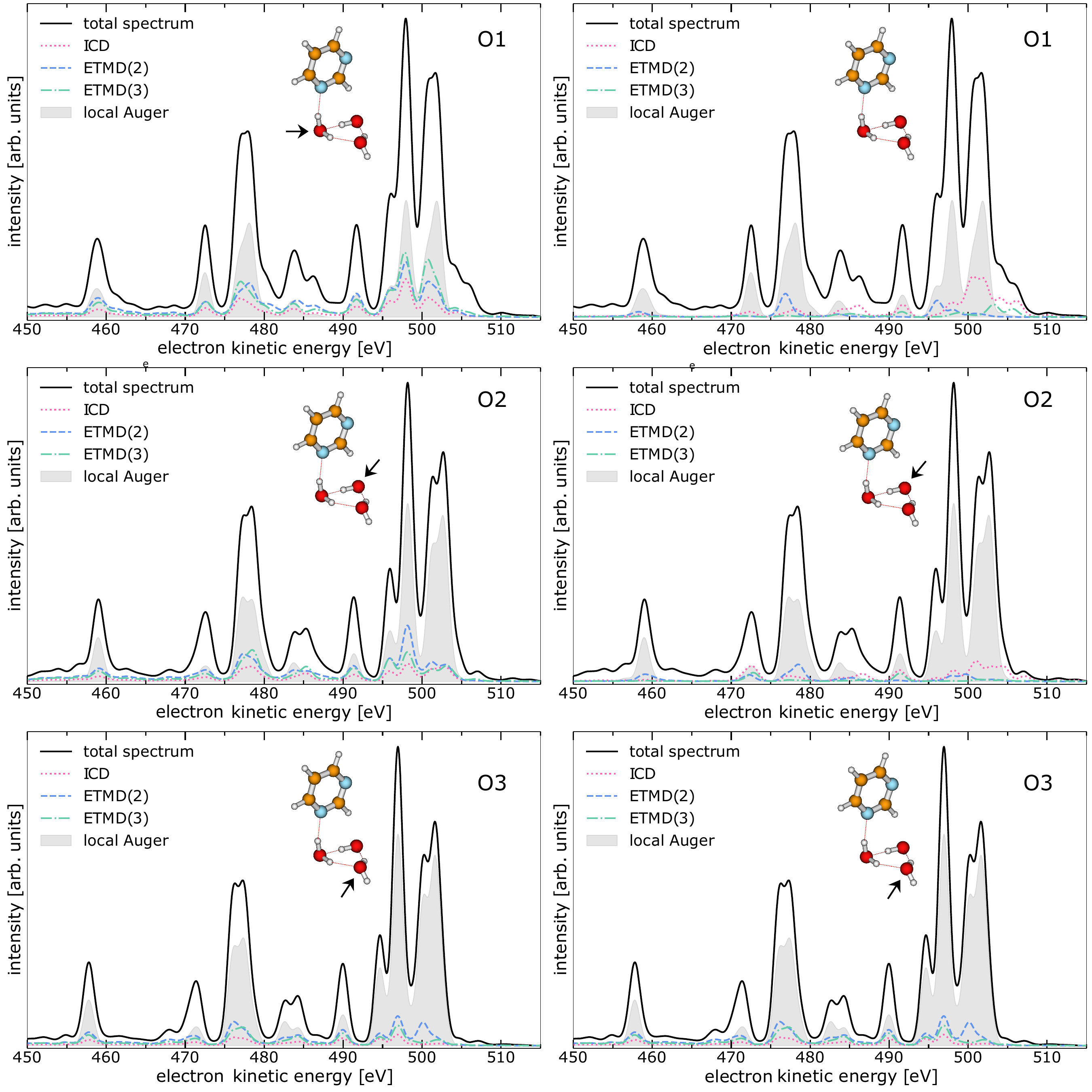}
  \caption{Theoretical spectra of the emitted electrons in the decay of an O\,1s vacancy in a pyrimidine-water cluster, consisting of one pyrimidine molecule and three water molecules.  The equilibrium geometry of the cluster is illustrated as insets. The different local  (Auger decay - gray filled traces) and intermolecular contributions (core-level ICD -  pink dotted traces, core-level ETMD(2) -  blue dashed traces, and ETMD(3) - green dashed-dotted traces)  of the  total spectra (black solid  traces) are presented. The spectra on the left side show the intermolecular processes with holes at the pyrimidine in the final states and the spectra  on the right side the intermolecular processes with holes only at water molecules in the finals states. Note that  identical total and Auger spectra are shown in all figures for comparison. The different rows depict spectra resulting from the ionization of different waters (marked with black arrow).}
  \label{fig:P-3W}
\end{figure}

\renewcommand{\thefigure}{S4}
\begin{figure}[h!]
\includegraphics[width=0.9\textwidth]{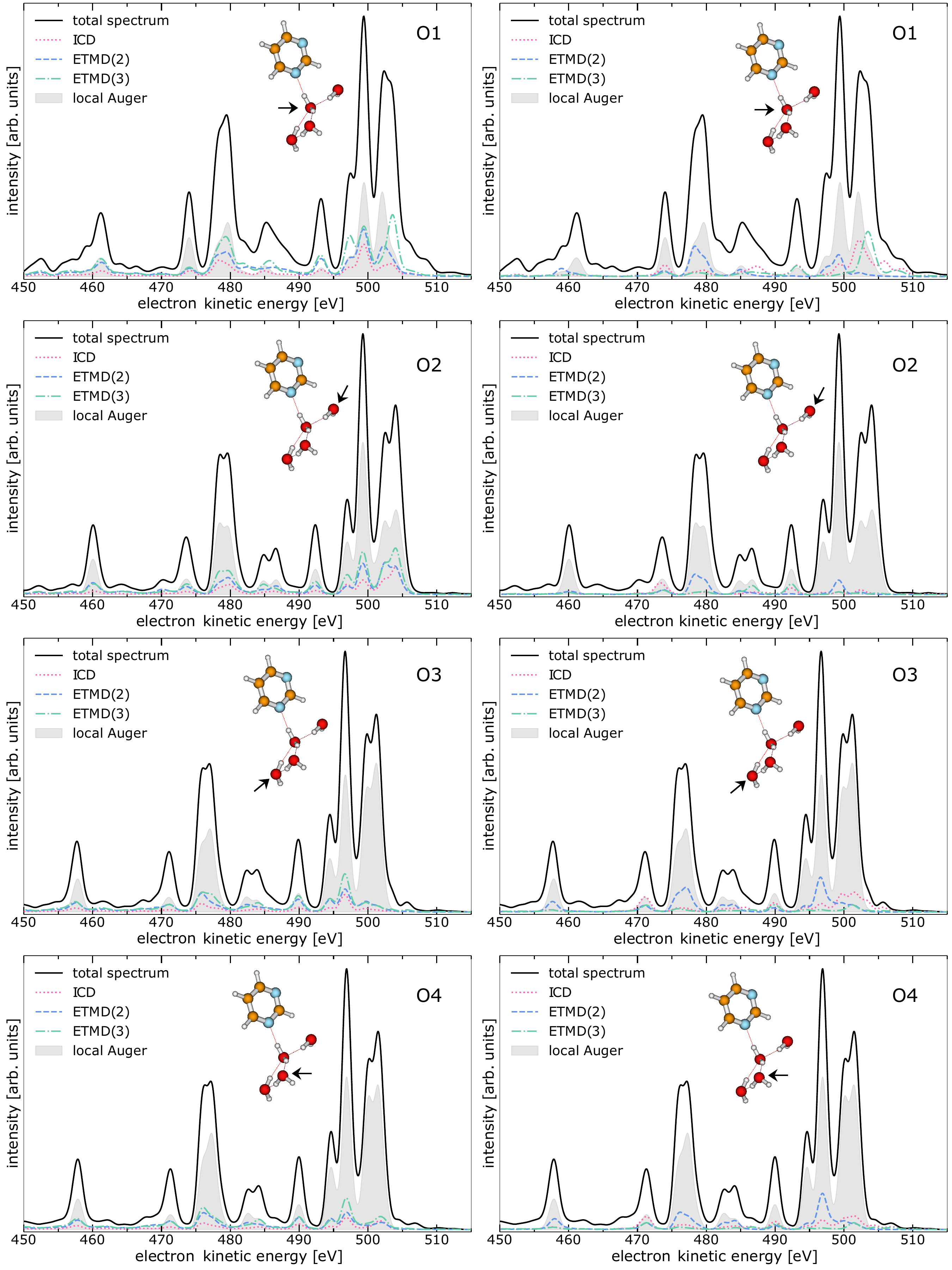}
\newpage
  \caption{Theoretical spectra of the emitted electrons in the decay of an O\,1s vacancy in a pyrimidine-water cluster, consisting of one pyrimidine molecule and four water molecules.  The equilibrium geometry of the cluster is illustrated as insets. The different local  (Auger decay - gray filled traces) and intermolecular contributions (core-level ICD -  pink dotted traces, core-level ETMD(2) -  blue dashed traces, and ETMD(3) - green dotted, dashed traces)  of the  total spectra (black solid  traces) are presented. The spectra on the left side show the intermolecular processes with holes at the pyrimidine in the final states and the spectra  on the right side the intermolecular processes with holes only at water molecules in the finals states. Note that  identical total and Auger spectra are shown in all figures for comparison. The different rows depict  spectra resulting from the ionization of different waters (marked with black arrow).}
  \label{fig:P-4W}
\end{figure}

\bibliography{library}
\bibliographystyle{naturemag}